\documentclass[twocolumn,a4paper,reprint,amssymb,aps,prd,groupedaddress,superscriptaddress,amsmath,nobibnotes,showpacs,nofootinbib]{revtex4-2}
\usepackage{graphicx,bm,color,psfrag}
\usepackage{amsfonts}
\usepackage{lipsum}
\usepackage{enumitem}
\usepackage[colorlinks = true,
            linkcolor = blue,
            urlcolor  = blue,
            citecolor = blue,
            anchorcolor = blue]{hyperref}

\usepackage{xcolor}
\usepackage{threeparttable,booktabs}

\newcommand{\be}{\begin{equation}}
\newcommand{\ee}{\end{equation}}

\begin{document}

\title{Cosmological models in scale-independent energy-momentum squared gravity}


\author{\"{O}zg\"{u}r Akarsu}
\email{akarsuo@itu.edu.tr}
\affiliation{Department of Physics, Istanbul Technical University, Maslak 34469 Istanbul, Turkey}

\author{N. Merve Uzun}
\email{nebiye.uzun@boun.edu.tr}
\affiliation{Department of Physics, Bo\u{g}azi\c{c}i University, Bebek 34342 Istanbul, Turkey}


\begin{abstract}

Scale-independent EMSG is a particular model of energy-momentum squared gravity (EMSG) in which the new terms in the Einstein field equations arising from the EMSG theory enter with the same power as the usual terms from Einstein-Hilbert part of the action. However, the model violates the local energy-momentum conservation and matter-current conservation in general and hence, permits a process of matter creation/annihilation in an expanding universe. Consequently, the scale factor dependencies of the energy densities are modified by the dimensionless model parameter $\alpha$. We revisit some nostalgias such as static universes and de Sitter/steady state universes. We reproduce the original ones, moreover, present some novelties, e.g., a spatially flat static universe, de Sitter expansion by negative vacuum energy, steady state universes in the presence of arbitrary fluids with constant equation of state (EoS) parameter other than dust, etc.  We also investigate the possible dynamics of dust dominated and radiation dominated universes. Depending on the value of $\alpha$, dust/radiation dominated universe exhibits power-law accelerated/decelerated expansion, corresponds to a steady state model or may end in a big rip. In the framework of anisotropic cosmology, we reproduce Barrow's quiescent universe in the presence of stiff fluid and extend it to fluids with arbitrary constant EoS parameter. We also relax the condition for isotropic initial singularity (big bang) owing to that EMSG effectively allows ultra-stiff EoS parameters.

\end{abstract}

\maketitle
\section{Introduction}
Subsequent to the establishment of general relativity (GR) in 1915, the first relativistic cosmological model was also put forward by Einstein himself in 1917~\cite{Einstein:1917ce}.  It was a static model having positively curved spatial sections with a uniform distribution of motionless matter on large scales. To achieve such a spatially finite but temporally infinite model, Einstein modified his theory by adding a cosmological constant that is positive ($\Lambda>0$) to the original field equations. His insistence on a static model was due to the fact that the prevailing opinion of his period was the Universe is unchanging in both space and time. Unfortunately, Einstein's static universe was proved to be unstable and eventually abandoned after the discovery of the expansion of Universe. The second cosmological model was de Sitter (dS) universe that is not static but stationary, devoid of any material source and contains only a positive cosmological constant [a negative cosmological constant produces anti-de Sitter (AdS) universe]~\cite{deSitter1917a,deSitter1917b}. It corresponds to a spatially flat universe expanding exponentially. This universe has no beginning or end, hence expands forever. However, such an exponentially expanding universe can be realized in the presence of matter with a constant energy density as well. This model proposed by Bondi, Gold, and Hoyle in 1948 was the steady state universe which requires continuous creation of matter in order to compensate for the diluting effect of the expansion on the matter energy density of the universe~\cite{Bondi:1948qk,Hoyle:1948zz}. To construct the steady state model, Hoyle modified the Einstein field equations (EFE) of GR by introducing a \textit{creation field} tensor, $C_{\mu\nu}$, which plays a role similar to that of the cosmological constant in the de Sitter model, and naturally, the matter creation process rendered the violation of the local/covariant energy-momentum conservation necessary. Although these three models of modern cosmology seem to possess different dynamics/ingredients, they were all inspired by the \textit{perfect cosmological principle}, which asserts that the Universe is homogeneous and isotropic (maximally symmetric) in space as well as homogeneous in time~\cite{Bondi:1948qk}.

 During the late 1940s and 1950s, the hot big bang along with the steady state model mentioned above were two major paradigms in cosmology. In contrast to the eternal universe of the steady state model, the hot big bang model suggests that the Universe should have once been started from an extremely hot and dense state, namely, an initial singularity~\cite{Gamow:1948pob}. Thus, the hot big bang violates the perfect cosmological principle, however, is in accordance with the so-called \textit{cosmological principle}, which states that the Universe is homogeneous and isotropic in space, but changes with time. After years of rivalry, the detection of the cosmic microwave background (CMB) -- expected relic radiation of an earlier hotter and denser phase of cosmic evolution -- in 1965 by Penzias and Wilson~\cite{Penzias:1965wn} favored hot big bang over steady state model. However, the CMB temperature at a value of about 2.7 K distributed isotropically over a wide range of angular scales gave rise to another question: How had the Universe from a generic initial singularity evolved to the isotropy we observe today? The first prominent attempt to explain the present isotropy was Misner’s chaotic cosmology~\cite{Misner:1967uu} which asserts that the Universe evolved from a generic, irregular and disordered state to today's large-scale order and regularity. According to the chaotic scenario, isotropy was generated by dissipative effects, such as particle creation or collisions and the present structure of the Universe is independent of the exact initial conditions at the big bang. Yet, it was shown that such a smoothing process may occur for limited type of anisotropies but not for generic ones~\cite{Stewart1969}. An alternative explanation, on the other hand, was Barrow's quiescent cosmology~\cite{Barrow1978} which suggests that the Universe indeed originated in a highly smooth and regular state in which a stiff fluid whose pressure equals to its energy density prevent the anisotropy from domination, and evolved towards irregularity through gravitational attraction. The quiescent universe scenario is also supported by Penrose’s ideas on gravitational entropy~\cite{Penrose1979}.

Gravity theories violating the local/covariant energy-momentum conservation can naturally result from modifying the form of the matter Lagrangian, $\mathcal{L}_{\rm m}$, in the Einstein-Hilbert action in a nonlinear way, for instance, by introducing some analytic function of the Lorentz scalar $T_{\mu \nu }T^{\mu \nu }$ constructed from the energy-momentum tensor (EMT), $T_{\mu \nu }$, of the material stresses \cite{Katirci:2014sti,Akarsu:2017ohj,Board:2017ign}, rather than generalizing the gravitational Lagrangian away from the linear function of scalar curvature, $R$, responsible for the Einstein tensor, $G_{\mu\nu}$. 
Such generalizations of GR result in new contributions by the usual material stresses to the right-hand side of the usual EFE without invoking any new type of sources and lead in general to nonconservation of the material stresses (see e.g.~\cite{Harko:2010mv,Harko:2011kv} for other similar type of theories). Of course, one may search for a curvature-type modification on the left-hand side of the usual EFE corresponding to a matter-type modification arising on the right-hand side of the usual EFE, but this might not be trivial or even possible and thereby such matter-type of modified gravity theories are quite rich and promising in establishing novel astrophysical and cosmological models.

 A particular example of this type of generalization dubbed \textit{energy-momentum squared gravity} (EMSG) has recently been proposed by the inclusion of the term $f(T_{\mu \nu }T^{\mu \nu })$ in the usual Einstein-Hilbert action~\cite{Katirci:2014sti,Akarsu:2017ohj,Board:2017ign}.  Some specific EMSG models studied in various contexts are  \textit{quadratic EMSG} represented by $f(T_{\mu \nu }T^{\mu \nu }) \propto T_{\mu \nu }T^{\mu \nu }$~\cite{Roshan:2016mbt}, \textit{energy-momentum powered gravity} (EMPG) represented by $f(T_{\mu \nu }T^{\mu \nu }) \propto (T_{\mu \nu }T^{\mu \nu })^{\eta}$~\cite{Akarsu:2017ohj,Board:2017ign} and \textit{energy-momentum log gravity} (EMLG) represented by $f(T_{\mu \nu }T^{\mu \nu }) \propto \ln{(T_{\mu \nu }T^{\mu \nu })}$~\cite{Akarsu:2019ygx}. These theories have been investigated in the literature from different motivations and perspectives~\cite{Katirci:2014sti,Akarsu:2017ohj,Board:2017ign,Roshan:2016mbt,Akarsu:2018zxl,Nari:2018aqs,Akarsu:2019ygx,Faria:2019ejh,Bahamonde:2019urw,Chen:2019dip,Barbar:2019rfn,Kazemi:2020hep,Singh:2020bdv,Nazari:2020gnu,Akarsu:2020vii,Chen:2021cts,Nazari:2022xhv,Nazari:2022fbn,Acquaviva:2022bju,Khodadi:2022zyz,Akarsu:2022abd,Akarsu:2018aro,Rudra:2020rhs,Khodadi:2022xtl,Tangphati:2022acb}. Some interesting features of the EMSG theory studied in the framework of these models are the allowing an EMT having a nonvanishing divergence~\cite{Akarsu:2017ohj,Board:2017ign}; the replacing of the initial singularity with an initial bounce through the higher order contributions of the energy density of dust~\cite{Roshan:2016mbt,Akarsu:2018zxl}; the possibility of driving late time accelerated expansion from the usual material sources without invoking a cosmological constant $\Lambda$~\cite{Akarsu:2017ohj}; the screening of the cosmological constant in the past by the altered scale factor dependency of dust due to the nonconservation of EMT~\cite{Akarsu:2019ygx}; the effective source that yields constant inertial mass density arising in~\cite{Akarsu:2019ygx}; the screening of the expansion anisotropy via the quadratic contribution of dust in the Friedmann equation and leading to mathematically exactly the same Friedmann equation with GR even in the presence of  anisotropic expansion~\cite{Akarsu:2020vii}; the altered past or far future of the Universe~\cite{Roshan:2016mbt,Akarsu:2018zxl,Acquaviva:2022bju}; etc.

In this paper, we investigate a particular model, initially proposed in Ref.~\cite{Akarsu:2018aro}, called \textit{scale-independent EMSG}  described by $f(T_{\mu \nu }T^{\mu \nu }) \propto  \sqrt{T_{\mu \nu }T^{\mu \nu }}$ (corresponding to the particular case $\eta=1/2$ of EMPG~\cite{Akarsu:2017ohj,Board:2017ign}). It is particular in the sense that the new contributions to the EFE of GR due to the EMSG modification enter with the same power that of the usual terms from Einstein-Hilbert part of the action, and thereby this EMSG modification affects the field equations independently of the energy density scale considered. On the other hand, the local energy-momentum conservation and the matter-current conservation are violated in general, hence the scale factor dependency of the standard energy density is altered. The violation of the matter-current conservation means that this model permits matter creation process and provides us with the opportunity to construct a steady state universe. The scale-independent EMSG has been studied so far as an extension of the Lambda cold dark matter ($\Lambda$CDM) model -- the current standard model of cosmology -- in which sources with different equations of state couple to the spacetime differently~\cite{Akarsu:2018aro}. In what follows, we investigate the cosmological models in scale-independent EMSG on theoretical ground. We revisit some nostalgias such as static and steady state universe models. We discuss the possible dynamics of dust/radiation dominated universes. Moreover, we present some novel scenarios that cannot be achieved in GR under same conditions regarding big rip and isotropic initial singularity. It is known that nowadays $\Lambda$CDM suffers from persistent tensions of various degrees of significance among some existing datasets~\cite{DiValentino:2020vhf,DiValentino:2020zio,DiValentino:2020vvd,DiValentino:2020srs,DiValentino:2021izs,Perivolaropoulos:2021jda,Abdalla:2022yfr}, thereby, we think that on the road to resolving the current issues and problems of cosmology, such historical considerations might enable us to develop new approaches that are hard to come to mind when we stay within the framework of the standard cosmological model.

\section{Scale-independent EMSG}

The action of EMSG is constructed by the addition of the term $f(T_{\mu\nu}T^{\mu\nu})$ to the Einstein-Hilbert action~\cite{Katirci:2014sti,Akarsu:2017ohj,Board:2017ign};
\begin{align}
S=\int \left[\frac{1}{2\kappa} R+f(T_{\mu\nu}T^{\mu\nu})+\mathcal{L}_{\rm m}\right]\sqrt{-g}\,{\rm d}^4x,
\label{action}
\end{align}
where $\kappa=8 \pi G$ is Newton's constant (scaled by $8\pi$), $R=g^{\mu\nu} R_{\mu\nu}$ is the Ricci scalar calculated from the Ricci tensor $R_{\mu\nu}$, $g$ is the determinant of the metric tensor $g_{\mu\nu}$, $\mathcal{L}_{\rm m}$ is the Lagrangian density corresponding to the matter source described by the energy-momentum tensor (EMT) $T_{\mu\nu}$, and units are used such that $c=1$. 

We vary the action above with respect to the inverse metric $g^{\mu\nu}$ as
\begin{equation}
 \begin{aligned}   \label{variation}
  \delta S=\int\, {\rm d}^4 x \bigg[&\frac{1}{2 \kappa}\frac{\delta (\sqrt{-g}R)}{\delta g^{\mu\nu}}+\frac{\delta(\sqrt{-g}\mathcal{L}_{\rm m})}{\delta g^{\mu\nu}}      \\
 &+f(T_{\sigma\epsilon}T^{\sigma\epsilon}) \: \frac{\delta(\sqrt{-g})}{\delta g^{\mu\nu}} \\
 &+\frac{\partial f}{\partial(T_{\lambda\xi}T^{\lambda\xi})}\frac{\delta(T_{\sigma\epsilon}T^{\sigma\epsilon})}{\delta g^{\mu\nu}}\sqrt{-g}\bigg] \delta g^{\mu\nu},  
\end{aligned}
\end{equation}
and, as usual, define the EMT in terms of the matter Lagrangian $\mathcal{L}_{\rm m}$ as
  \begin{equation}
  \label{tmunudef}
 T_{\mu\nu}=-\frac{2}{\sqrt{-g}}\frac{\delta(\sqrt{-g}\mathcal{L}_{\rm m})}{\delta g^{\mu\nu}}=g_{\mu\nu}\mathcal{L}_{\rm m}-2\frac{\partial \mathcal{L}_{\rm m}}{\partial g^{\mu\nu}},
 \end{equation}
for which we assume $\mathcal{L}_{\rm m}$ depends only on the metric tensor components, and not on its derivatives since this is the case for the Maxwell field and gauge fields in general, as well as for scalar fields. Consequently, the modified Einstein field equations read
\begin{equation}
G_{\mu\nu} =\kappa  \left[ T_{\mu\nu}+fg_{\mu\nu}-2\frac{\partial f}{\partial(T_{\sigma\epsilon}T^{\sigma\epsilon})}\theta_{\mu\nu}\right],
\label{modfieldeq}
\end{equation}
where $G_{\mu\nu}=R_{\mu\nu}-\frac{1}{2}Rg_{\mu\nu}$ is the Einstein tensor; $\theta_{\mu\nu}$ is a new tensor defined as
\begin{equation}
\begin{aligned}
\theta_{\mu\nu}=  \frac{\delta (T_{\sigma\epsilon}T^{\sigma\epsilon})}{\delta g^{\mu\nu}}&=-2\mathcal{L}_{\rm m}\left(T_{\mu\nu}-\frac{1}{2}g_{\mu\nu}T\right)-T T_{\mu\nu}\\
&\quad\,+2T_{\mu}^{\lambda}T_{\nu\lambda}-4T^{\sigma\epsilon}\frac{\partial^2 \mathcal{L}_{\rm m}}{\partial g^{\mu\nu} \partial g^{\sigma\epsilon}},
\label{theta}
\end{aligned}
\end{equation}
where $T$ is the trace of the EMT, $T_{\mu\nu}$. In the present study, following the literature to date on EMSG and on similar theories (see~\cite{Harko:2011kv,Haghani:2013oma,Odintsov:2013iba}), we consider $\mathcal{L}_{\rm m}=p$ which gives rise to the perfect fluid EMT given in Eq.~\eqref{em} through the definition~\eqref{tmunudef}. Accordingly, we take the advantage of setting the last term of the tensor $\theta_{\mu\nu}$ given in Eq.~\eqref{theta} to zero, namely, $\partial^2 \mathcal{L}_{\rm m}/(\partial g^{\mu\nu} \partial g^{\sigma\epsilon})=0$. In this way, we are able to calculate $\theta_{\mu\nu}$ independently of the function $f$~\cite{Board:2017ign}.

We proceed with the \textit{scale-independent EMSG} \cite{Akarsu:2018aro} described by
\begin{equation}
\label{eqn:powerassumption}
f(T_{\mu\nu}T^{\mu\nu})= \alpha \sqrt{T_{\mu\nu}T^{\mu\nu}},
\end{equation}
and accordingly, the action in \eqref{action} becomes
\begin{equation}
S=\int \left[\frac{1}{2 \kappa}R+\alpha \sqrt{T_{\mu\nu}T^{\mu\nu}}+\mathcal{L}_{\rm m}\right]\sqrt{-g}\,{\rm d}^4x,
\label{eq:action}
\end{equation}
where $\alpha$ is a dimensionless constant that determines the gravitational coupling strength of the scale-independent EMSG modification to GR. Accordingly, the modified  Einstein field equations~\eqref{modfieldeq} for the action~\eqref{eq:action} now read,
\begin{equation}
G_{\mu\nu}=\kappa T_{\mu\nu}+\kappa \tilde T_{\mu\nu},
\label{fieldeq}
\end{equation}
where
\begin{equation}   \label{emteff}
 \tilde T_{\mu\nu}=\alpha \sqrt{T_{\sigma\epsilon}T^{\sigma\epsilon}}\bigg(g_{\mu\nu}- \frac{\theta_{\mu\nu}}{T_{\lambda\xi}T^{\lambda\xi}}\bigg),
\end{equation}
describes the effective source due to the EMSG modification stated in Eq.~\eqref{eqn:powerassumption}. From \eqref{fieldeq}, the covariant divergence of the EMT reads $\nabla^{\mu}T_{\mu\nu}=-\nabla^{\mu}\tilde T_{\mu\nu}$ which can be explicitly written as
\begin{equation}
\begin{aligned}
  \label{nonconservedenergy}
\nabla^{\mu}T_{\mu\nu}=-\alpha g_{\mu\nu}\nabla^{\mu}\sqrt{T_{\sigma\epsilon}T^{\sigma\epsilon}}+\alpha \nabla^{\mu}\left[\frac{\theta_{\mu\nu}}{\sqrt{T_{\sigma\epsilon}T^{\sigma\epsilon}}}\right].
\end{aligned}
\end{equation}
We note that, unless $\alpha=0$ (which guarantees $\tilde T_{\mu\nu}=0$), the right-hand side of this equation does not vanish in general, and thus the EMT is not necessarily conserved, i.e., $\nabla^{\mu}T_{\mu\nu}=0$ is not necessarily satisfied.

We consider the perfect fluid form of the EMT;
\begin{align}
\label{em}
T_{\mu\nu}=(\rho+p)u_{\mu}u_{\nu}+p g_{\mu\nu},
\end{align} 
where $\rho>0$ (unless otherwise stated) is the energy density and $p$ is the thermodynamic pressure of the fluid; $u_{\mu}$ is the four-velocity satisfying the conditions $u_{\mu}u^{\mu}=-1$ and $\nabla_{\nu}u^{\mu}u_{\mu}=0$. For the perfect fluid above, the self-contraction of EMT and the tensor $\theta_{\mu\nu}$ read
\begin{align}
T_{\sigma\epsilon}T^{\sigma\epsilon}&= \rho^2+3p^2, \label{perfT2} \\
\theta_{\mu\nu}&=-  (\rho+p) (\rho+3p) u_{\mu} u_{\nu}.  \label{perftheta}
\end{align}
Substituting \eqref{em} along with \eqref{perfT2} and \eqref{perftheta} into the effective EMT defined in \eqref{emteff}, we obtain
\begin{equation}    \label{eqn:emsgtmunu}
\begin{aligned}
  \tilde T_{\mu\nu}=&\alpha\bigg(\sqrt{\rho^{2}+3p^2}+\frac{4 \rho p}{\sqrt{\rho^{2}+3p^2}}\bigg)u_{\mu} u_{\nu}\\
  &+\alpha\sqrt{\rho^{2}+3p^2}\,g_{\mu\nu} ,
  \end{aligned}
\end{equation}
which, comparing with the EMT of the form that of perfect fluid, i.e., $\tilde T_{\mu\nu}=(\tilde \rho+\tilde p)u_{\mu}u_{\nu}+ \tilde p g_{\mu\nu}$, implies
\begin{align}
\tilde \rho=\frac{4 \alpha \rho p}{\sqrt{\rho^{2}+3p^2}}\quad\textnormal{and}\quad  
\tilde p=\alpha \sqrt{\rho^{2}+3p^2},
\end{align}
which are the energy density and pressure of the effective source, respectively. Thus, defining a total EMT as $T_{\mu\nu}^{\rm tot}= T_{\mu\nu}+\tilde T_{\mu\nu}=(\rho_{\rm tot}+ p_{\rm tot})u_{\mu}u_{\nu}+ p_{\rm tot} g_{\mu\nu}$, the total energy density and pressure can be identified as follows:
\begin{equation}
    \rho_{\rm tot}\equiv\rho+\tilde\rho\quad\textnormal{and}\quad p_{\rm tot}\equiv p+\tilde p.
\end{equation}
Considering barotropic fluid described by the equation of state (EoS) of the form $ p=w\rho$, where $w$ is the EoS parameter we assume to be constant (here, and henceforth), we obtain
\begin{align}
\rho_{\rm tot}&=\rho\left(1+\frac{4 \alpha w}{\sqrt{1+3w^2}}\right), \label{effdens} \\
p_{\rm tot}&=\rho\left(w+\alpha \sqrt{1+3w^2}\right). \label{effpres}
\end{align}
 We use Eqs.~\eqref{effdens} and~\eqref{effpres} in the total EMT, and from the vanishing divergence of it, that is, $\nabla_{\mu} T^{\mu\nu}_{\rm tot}=0$, we obtain the matter-current conservation as follows:
 \begin{equation}
     \nabla_{\mu}(\rho u^{\mu})= - \rho \left[w+\frac{\alpha (1-w^2)}{4 \alpha w+\sqrt{1+3w^2}}\right] \nabla_{\mu} u^{\mu}.
 \end{equation}
 This reduces to $\nabla_{\mu}(\rho u^{\mu})= -\alpha\rho\nabla_{\mu} u^{\mu}$ for dust ($w=0$), implying that baryon number density is not necessarily conserved. Here $\nabla_{\mu}u^{\mu} \equiv \Theta$ is the volume expansion rate of the fluid---in case of Robertson-Walker (RW) spacetime metric, we have $\Theta=3 H$, where $H=\frac{\dot{a}}{a}$ is the Hubble parameter with $a$ being the scale factor and a dot denotes $\frac{{\rm d}}{{\rm d}t}$ (see Eq.~\eqref{rw}). Namely, the term with $\alpha\neq0$ on the right-hand side shows us that in an expanding universe ($\Theta>0$) governed by EMSG, there is continuous creation (for $\alpha<0$) and annihilation (for $\alpha>0$) of matter on cosmological scales, namely, in the expanding space between galaxies, but not within galaxies---the galaxies (each is a gravitationally bound system) themselves are independent of the expansion of the Universe, meaning they do not expand (i.e., $\Theta=0$ in the local region of the Universe occupied by a galaxy). This property of the model signals the possibility of constructing a steady state model in the presence of a material source.

   We proceed by considering the spatially maximally symmetric spacetime metric, i.e., the RW spacetime metric, with flat spacelike sections given as
\begin{equation}
{\rm d}s^{2}=-{\rm d}t^{2}+a^{2}({\rm d}x^{2}+{\rm d}y^{2}+{\rm d}z^{2}),  \label{rw}
\end{equation}
where the scale factor $a = a(t)$ is a function of cosmic time $t$ only. Therefore, the modified Friedmann equations read
\begin{align}   \label{fieldeqn}
    \frac{{\dot a}^2}{a^2}=\frac{\kappa}{3}\rho_{\rm tot} \quad ,\quad
    \frac{\ddot a}{a}=-\frac{\kappa}{6}(\rho_{\rm tot}+3 p_{\rm tot}),
\end{align}
or equivalently, in an explicit form as follows:
\begin{align}  
    \frac{{\dot a}^2}{a^2}=\frac{\kappa}{3}\rho&\left(1+\frac{4 \alpha w}{\sqrt{1+3w^2}}\right),\label{eqn:1FE}\\
     \frac{\ddot a}{a}=-\frac{\kappa}{6}\rho&\left(1+3w+\alpha\frac{3+4w+9w^2}{\sqrt{1+3w^2}}\right).\label{eqn:2FE}
\end{align}
The corresponding continuity equation can be written as
\begin{equation}  \label{conteqn}
    \dot \rho+3\frac{\dot a}{a}\rho(1+w)=-3\frac{\dot a}{a}\rho\frac{\alpha (1-w^2)}{4\alpha w+\sqrt{1+3w^2}}.
\end{equation}
The expression on the right-hand side is the modification arising from EMSG and is equal to zero in the case $\alpha=0$, corresponding to GR. We realize that this term also drops for $w=-1$ and $w=1$, which then would give the same scale factor dependencies for the vacuum energy ($\rho=\rm const$) and stiff fluid ($\rho \propto a^{-6}$) as in GR. On the other hand, the model, in general, modifies the evolutions of the energy densities, including those of the known sources such as dust ($w=0$) and radiation ($w=\frac{1}{3}$), as 
\begin{equation}
 \rho \propto a^{-3 (1+w)\left[1+\frac{\alpha(1-w)}{4\alpha w+\sqrt{1+3w^2}}\right]} .  \end{equation}

In this paper, we restrict our study to positive-definite values of the total energy density. Therefore, from Eq.~\eqref{effdens}, we deduce that, if $\rho>0$ is imposed and $w\neq0$, then the positivity of total energy density, $\rho_{\rm tot} > 0$, requires
\begin{equation}
  \begin{aligned}
    \alpha < -\frac{\sqrt{1+3w^2}}{4w}\quad\textnormal{for}\quad w<0,\\
    \alpha > -\frac{\sqrt{1+3w^2}}{4w}\quad\textnormal{for}\quad w>0,
\end{aligned}
\end{equation}
whereas $\alpha$ can take arbitrary values when $w=0$, which describes dust. In particular, $\rho_{\rm tot} > 0$ implies $\alpha < \frac{1}{2}$ for positive ($\alpha > \frac{1}{2}$ for negative) vacuum energy described by $w=-1$, $\alpha > -\frac{\sqrt{3}}{2}$ for the radiation described by $w=\frac{1}{3}$, and $\alpha > -\frac{1}{2}$ for the stiff fluid described by $w=1$.

\begin{figure}[t!]
\centering
\includegraphics[width=1\linewidth]{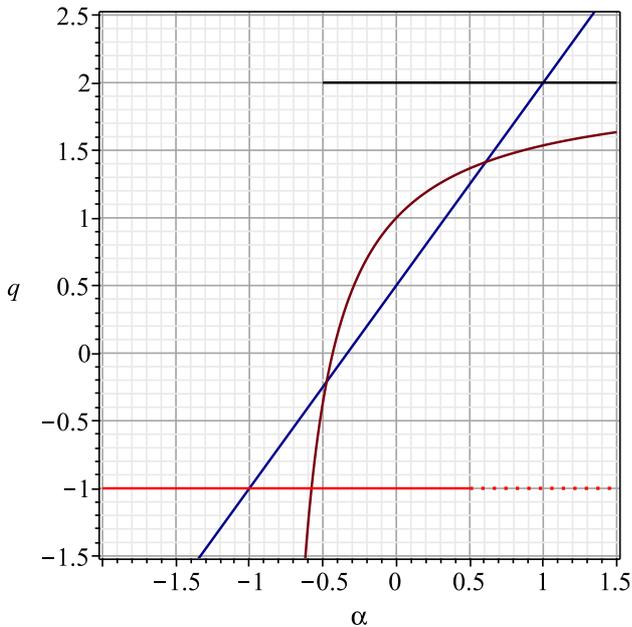} 
\caption{Deceleration parameter $q$ versus $\alpha$ graph. Plotted for stiff fluid ($w=1$, black), radiation ($w=\frac{1}{3}$, brown), dust ($w=0$, blue), and positive vacuum energy ($w=-1$, red).}
\label{fig:comparison}
\end{figure}

Next, we calculate the deceleration parameter, $q \equiv   -\frac{{\ddot a}a}{{\dot a}^2}$, as follows:
\begin{equation}
    q=\frac{1}{2}+\frac{3}{2}\left[w+\frac{\alpha (1-w^2)}{4 \alpha w+\sqrt{1+3w^2}}\right],
\end{equation}
which reduces to $q=\frac{1}{2}+\frac{3}{2}w$ in GR ($\alpha=0$). For arbitrarily large values of $\alpha$, i.e, $\alpha\rightarrow\infty$, the deceleration parameter reads $q\rightarrow\frac{1}{2}+\frac{9}{8}w+\frac{3}{8}w^{-1}$. We display the deceleration parameter curves of different sources in Fig.~\ref{fig:comparison}. We note that for two particular values of the parameter $\alpha$, the value of the deceleration parameter does not differ between the radiation (r) and dust (d); $q_{\rm r}=q_{\rm d}=1.41$ for $\alpha=0.61$ and $q_{\rm r}=q_{\rm d}=-0.21$ for $\alpha=-0.47$. The model exhibits de Sitter expansion ($q=-1$) at $\alpha=-1$ for dust and at $\alpha=-\frac{\sqrt{3}}{3}$ for radiation. In the case of vacuum energy (positive or negative) and stiff fluid, the values of $q$ do not differ from the ones in GR; that is, $q=-1$ for $w=-1$ and $q=2$ for $w=1$. Interestingly, in the case of radiation, for arbitrarily large values of $\alpha$ the value of the deceleration parameter approaches two, i.e., $q_{\rm r}\rightarrow2$ as $\alpha\rightarrow\infty$. We have super exponential expansion ($q<-1$), viz., big rip, for $-\frac{\sqrt{3}}{2}<\alpha<-\frac{\sqrt{3}}{3}$ (where the lower boundary corresponds to $\rho_{\rm tot}>0$) in the presence of radiation and for $\alpha<-1$ in the presence of dust.

\section{Static Universes}
The simplest relativistic cosmological model satisfying the perfect cosmological principle, which states that the Universe is homogeneous and isotropic in space as well as homogeneous in time~\cite{Bondi:1948qk}, in the presence of a material source with a positive energy density is the static universe. For such a universe having a constant scale factor ($a=\rm const$), it is required that $\dot{a}=0$ and $\ddot{a}=0$, which imply $\rho_{\rm tot}=0$ and $p_{\rm tot}=0$ according to the modified Friedmann equations \eqref{fieldeqn} of EMSG. After using Eqs.~\eqref{effdens} and~\eqref{effpres} it turns out that there are two cases satisfying these conditions: (i)  In the presence of positive vacuum energy ($\rho>0$ with $w=-1$) provided that $\alpha=\frac{1}{2}$, whereas this source would lead to a de Sitter universe in GR ($\alpha=0$). (ii)  In the presence of stiff fluid with a positive energy density ($\rho>0$ with $w=1$) provided that $\alpha=-\frac{1}{2}$, whereas this source would lead to $a\propto t^\frac{1}{3}$ in GR ($\alpha=0$).

 Historically, on the other hand, the first such relativistic cosmological model is the Einstein's static universe (aka the Einstein universe or the Einstein static eternal universe) filled with dust and finite with a positive (spherical) curvature \cite{Einstein:1917ce}. To make a comparison, we now include the spatial curvature in our model. Substituting $\frac{k}{a^2}$ term on the left-hand side of Eq.~\eqref{eqn:1FE} and using along with Eq.~\eqref{eqn:2FE}, a static universe having a positive spatial curvature ($k=1$) can be obtained with a $w$-dependent coupling parameter
 \begin{equation}
   \alpha =-\frac{(1+3 w) \sqrt{1+3 w^2}}{3+4w+9 w^2} ,
 \end{equation} 
and the constant positive energy density
\begin{equation}
 \rho =\frac{ \left(3+4w+9 w^2\right)}{\kappa a^2 \left(1-w^2\right)},
 \end{equation}
for $-1<w<1$. In the case of dust, we have $\alpha = -\frac{1}{3}$ and $a=\sqrt{\frac{3}{\kappa \rho_{\rm d}}}$. Thus, we find that a dust-filled static universe with a positive spatial curvature is possible without the need of introducing a cosmological constant in contrast to GR. It is also noteworthy that in the scale-independent EMSG, any material source with a constant EoS parameter can generate a static universe that is spatially flat in the presence of a fluid described by $w=\pm1$ and positively curved in the presence of a fluid with $-1<w<1$.
 
\section{de Sitter Universes}
Another model that satisfies the perfect cosmological principle is the standard de Sitter universe which is an expanding model through a positive cosmological constant devoid of any material source (a negative cosmological constant induces anti-de Sitter universe)~\cite{deSitter1917a,deSitter1917b}. Also, a universe expanding in such a way can be realized in the presence of a material source with constant energy density as well, and is named steady state universe~\cite{Bondi:1948qk,Hoyle:1948zz}.

To obtain a de Sitter universe, viz., an exponentially expanding universe, we choose
\begin{equation}
    \rho_{\rm tot}=\textnormal{const}>0,
\end{equation}
which, using in the modified Friedmann equations \eqref{fieldeqn}, leads to
\begin{equation}
    a\propto e^{\sqrt{\frac{\kappa}{3}\rho_{\rm tot}} \,t}\quad\textnormal{and}\quad p_{\rm tot}=-\rho_{\rm tot},    
\end{equation} 
like in the standard de Sitter solution in GR provided by the usual positive vacuum energy, which yields $\rho_{\rm vac}=\rm const>0$ with $p_{\rm vac}=-\rho_{\rm vac}$ (i.e., $w_{\rm vac}=-1$). We remark that the standard de Sitter universe in GR is empty, namely, it does not contain material sources (hence, there is no observer in the universe). However, in the scale-independent EMSG, the de Sitter universe can be realized in the presence of any material source (hence, there can be observer in the universe), provided that it has a constant EoS parameter, $w=\rm const$, for which two cases,  (i) $w=-1$, and (ii) $w\neq-1$, should be investigated separately.

\subsection{de Sitter Universe from dS/AdS vacua ($w=-1$)}
In the presence of vacuum energy ($w=-1$), which can yield either positive or negative constant energy density values, $\rho_{\rm vac}$, the total EMT reads $T_{\mu\nu}^{\rm tot}=-(1-2\alpha)\rho_{\rm vac} g_{\mu\nu}$; resembling the usual vacuum energy with an energy density scaled by $(1-2\alpha)$ as follows;
\begin{equation}
   \rho_{\rm tot}=(1-2\alpha)\rho_{\rm vac}.
\end{equation}
The condition $\rho_{\rm tot}>0$ implies (i) $\alpha<\frac{1}{2}$ for de Sitter (dS) vacua ($\rho_{\rm vac}>0$), and (ii) $\alpha>\frac{1}{2}$ for anti-de Sitter (AdS) vacua ($\rho_{\rm vac}<0$). The latter is of particular interest as in this case, i.e., when $\alpha>\frac{1}{2}$, de Sitter expansion, which is by positive valued vacuum energy density in GR, is driven by negative vacuum energy density in the scale-independent EMSG. A negative vacuum energy density is not only ubiquitous in the fundamental theoretical physics without any complication, but also a theoretical sweet spot; the negative vacuum energy density (produces an AdS background in GR and many other gravity theories, but a dS background in the scale-independent EMSG) is welcome due to the celebrated AdS/CFT (conformal field theory) correspondence~\cite{Maldacena:1997re} and is preferred by string theory and string-theory-motivated supergravities~\cite{Bousso:2000xa}. In contrast, the positive vacuum energy density suffers from theoretical challenges: getting a vacuum solution with a positive cosmological constant within string theory or formulating QFT on the background of a dS space (provided by $\rho_{\rm vac}>0$) has been a notoriously difficult task.~\cite{Weinberg:1988cp,Sahni:1999gb}

\subsection{Steady State Universe: de Sitter universe in the presence of any material source ($w\neq-1$)}   \label{sec:ssu}
In the presence of a source with a constant EoS parameter different than minus unity, if we impose $\rho= \rm const$, it turns out that the parameter $\alpha$, the gravitational coupling strength of the EMSG modification to GR, is not arbitrary;
\begin{equation}
    \alpha=-\frac{\sqrt{1+3w^2}}{1+3w},
\end{equation}
 which can be seen from Eq.~\eqref{conteqn}. The total energy density~\eqref{effdens} is also constant in this case and reads,
\begin{equation}
   \rho_{\rm tot}=\rho\left(1-\frac{4w}{1+3w}\right),
\end{equation}
in which, due to the condition $\rho_{\rm tot}>0$, the EoS parameter must satisfy $-\frac{1}{3}<w<1$. We have $\alpha=-1$ and $\rho_{\rm tot}=\rho_{\rm d}$ for $w=0$ (dust), and $\alpha=-\frac{\sqrt{3}}{3}$ and $\rho_{\rm tot}=\frac{\rho_{\rm r}}{3}$ for $w=\frac{1}{3}$ (radiation). We note that the effective source due to the scale-independent EMSG in this case resembles the usual vacuum energy, whose density value is constant and proportional to that of the physical source (which also yields constant energy density $\rho=\rm const$), but is scaled by the EoS parameter of the source;
\begin{equation}
    \tilde \rho=-\frac{4w}{1+3w}\rho.
\end{equation}
Due to the fact that the energy density of the source remains constant despite the expansion of the universe, this is a steady state universe model extending the original one~\cite{Hoyle:1948zz}, which assumes the only source in the universe is dust, to sources with arbitrary constant EoS parameter. In the presence of dust, we reproduce the original steady state universe model. Indeed, assuming dust, we find that $ \rho_{\rm d}={\rm const}$ and $p_{\rm d}=0$ while the effective source (for which $\alpha=-1$) reads $\tilde\rho_{\rm d}=0$ and $\tilde p_{\rm d}=-\rho_{\rm d}$. Substituting these into Eq.~\eqref{eqn:emsgtmunu} gives $\tilde T_{\mu\nu}=-\rho_{\rm d} (u_{\mu} u_{\nu}+g_{\mu\nu}),$ from which we have $\tilde T_{00}= 0$ and $\tilde T_{ij}=-\rho_{\rm d} g_{ij}$. 
We note that $\tilde T_{\mu\nu}^{\rm dust}$ is indeed equivalent to the \textit{creation field} tensor Hoyle~\cite{Hoyle:1948zz} used to modify the Einstein's field equations of GR, viz., $C_{\mu\nu}=\kappa\tilde T_{\mu\nu}^{\rm dust}$ (when $\alpha=-1$).

\section{Dust dominated universe}
In this section, we will consider a dust dominated universe in the framework of scale-independent EMSG. Assuming $\rho_{\rm d}>0$ and $p_{\rm d}=0$ (i.e., $w_{\rm d}=0$) in~Eq.~\eqref{eqn:emsgtmunu} gives rise to 
\begin{align}    \label{rhopeff}
\tilde T_{\mu\nu}^{\rm dust}= \alpha \rho_{\rm d} (u_{\mu} u_{\nu}+g_{\mu\nu}),
\end{align}
in which we have an effective source as
\begin{align}
\tilde\rho_{\rm d}=0\quad \textnormal{and}\quad
\tilde p_{\rm d}=\alpha\rho_{\rm d},\label{effpresdust}
\end{align}
which has nonzero pressure but a vanishing energy density, and hence does not correspond to a physical source. However, the dust itself behaves like a barotropic fluid with constant EoS parameter equal to $\alpha$.

The modified Friedmann equations~\eqref{eqn:1FE} and~\eqref{eqn:2FE} in the case of dust read
\begin{align}  \label{fe:dust}
    \frac{{\dot a}^2}{a^2}=\frac{\kappa}{3}\rho_{\rm d}\quad\textnormal{and}\quad
    \frac{\ddot a}{a}=-\frac{\kappa}{6}(1+3\alpha)\rho_{\rm d},
\end{align}
and the continuity equation~\eqref{conteqn} takes the following form
\begin{equation}  \label{conteqn-dust}
    \dot \rho_{\rm d}+3\frac{\dot a}{a}\rho_{\rm d}=-3\alpha\frac{\dot a}{a}\rho_{\rm d}.
\end{equation}
Let us now discuss the possible solutions of this model. We recall that there is no restriction on $\alpha$ stemming from the condition of $\rho_{\rm tot}>0$ and classify the solutions into three categories as follows:
\begin{itemize}
    \item The case $\alpha>-1$: In this case, the scale factor and energy density are
\begin{equation}
   a\propto t^{\frac{2}{3+3\alpha}}\quad\textnormal{and}\quad\rho_{\rm d} \propto a^{-3(1+\alpha)}.
\end{equation}
From the acceleration equation given in~\eqref{fe:dust}, one can see that we have decelerated expansion for $\alpha>-\frac{1}{3}$ and accelerated expansion for $-1<\alpha<-\frac{1}{3}$. When $\alpha=-\frac{1}{3}$, there is no acceleration/deceleration and the expansion is linear (i.e., $a \propto t$).
\item
The case $\alpha=-1$: This case implies $\dot{\rho}=0$ and gives rise to exponential expansion, therefore we obtain
\begin{equation}
   a\propto e^{\sqrt{\frac{\kappa\rho_{\rm d}}{3}}\,t}\quad\textnormal{and}\quad\rho_{\rm d}=\rm const.
\end{equation}
We note that this solution matches with the steady state universe (in the case of $w=0$) discussed in Section~\ref{sec:ssu}.
\item
 The case $\alpha<-1$: This case leads to a big rip, which is driven by a phantom field in GR~\cite{Caldwell:1999ew,Caldwell:2003vq}. We find
\begin{equation}
 a\propto(t_{\rm c}-t)^{\frac{2}{3+3\alpha}}\quad\textnormal{and}\quad\rho_{\rm d} \propto a^{-3(1+\alpha)},
\end{equation}
where $t\leq t_{\rm c}$ with $t_{\rm c}$ being the critical time at which $a\rightarrow\infty$. It is notable that we obtain big rip without introducing a phantom source ($w<-1$), but from dust ($w_{\rm d}=0$) itself.
\end{itemize}

\section{Radiation dominated universe}
Let us consider a radiation dominated universe in the same way. We assume $ \rho_{\rm r}>0$ and $p_{\rm r}=\frac{\rho_{\rm r}}{3}$ (i.e., $w_{\rm r}=\frac{1}{3}$). Then,~Eq.~\eqref{eqn:emsgtmunu} leads to
\begin{align}    \label{em-rad}
\tilde T_{\mu\nu}^{\rm rad}= \frac{4}{\sqrt{3}}\alpha\rho_{\rm r} u_{\mu} u_{\nu}+ \frac{2}{\sqrt{3}}\alpha\rho_{\rm r} g_{\mu\nu},
\end{align}
in which we have an effective source as
\begin{align}
\tilde\rho_{\rm r}=\tilde p_{\rm r}=\frac{2}{\sqrt{3}}\alpha\rho_{\rm r}\label{effpresrad}
\end{align}
that is reminiscent of a barotropic fluid with a constant EoS parameter equal to unity.

The modified Friedmann equations in the case of radiation read
\begin{align}   \label{fe:rad}
    \frac{{\dot a}^2}{a^2}=\frac{\kappa}{3}\left(1+\frac{2}{\sqrt{3}}\alpha\right)\rho_{\rm r}\quad\textnormal{and}\quad
    \frac{\ddot a}{a}=-\frac{\kappa}{3}\left(1+\frac{4}{\sqrt{3}}\alpha\right)\rho_{\rm r},
\end{align}
and the continuity equation~\eqref{conteqn} becomes
\begin{equation}  \label{conteqn-rad}
    \dot \rho_{\rm r}+4\frac{\dot a}{a}\rho_{\rm r}=-\frac{4 \alpha}{2 \alpha+\sqrt{3}}\frac{\dot a}{a}\rho_{\rm r}.
\end{equation}
We recall that the positivity of the total energy density, $\rho_{\rm tot}>0$, implies $\alpha>-\frac{\sqrt{3}}{2}$.
The Eq.~\eqref{conteqn-rad} splits the possible solutions into three categories as follows:
\begin{itemize}
    \item The case $\alpha>-\frac{\sqrt{3}}{3}$: In this case, the scale factor and energy density are
\begin{equation}
   a\propto t^{\frac{2\alpha+\sqrt{3}}{6\alpha+2\sqrt{3}}}\quad\textnormal{and}\quad\rho_{\rm r }\propto a^{-\frac{12\alpha+4\sqrt{3}}{2\alpha+\sqrt{3}}}.
\end{equation}
The acceleration equation given in~\eqref{fe:rad} implies that we have decelerated expansion for $\alpha>-\frac{\sqrt{3}}{4}$, and in particular $q_{\rm r}\rightarrow2$ for $\alpha\rightarrow\infty$ (see Fig.~\ref{fig:comparison}). In the range $-\frac{\sqrt{3}}{3}<\alpha<-\frac{\sqrt{3}}{4}$, we have power-law accelerated expansion. When $\alpha=-\frac{\sqrt{3}}{4}$, there is no acceleration/deceleration and the expansion is linear (i.e., $a \propto t$).
\item
The case $\alpha=-\frac{\sqrt{3}}{3}$: This case requires $\dot{\rho}=0$, and hence, leads to
\begin{equation}
   a\propto e^{\frac{\sqrt{\kappa \rho_{\rm r}}}{3}\,t}\quad\textnormal{and}\quad\rho_{\rm r}=\rm const.
\end{equation}
We note that this exponentially expanding  solution matches with the steady state universe (in the case of $w=\frac{1}{3}$) discussed in Section~\ref{sec:ssu}.
\item
The case $-\frac{\sqrt{3}}{2}<\alpha<-\frac{\sqrt{3}}{3}$: In this case, the universe ends in a big rip. We see that
\begin{equation}
   a\propto (t_{\rm c}-t)^{\frac{2\alpha+\sqrt{3}}{6\alpha+2\sqrt{3}}}\quad\textnormal{and}\quad\rho_{\rm r }\propto a^{-\frac{12\alpha+4\sqrt{3}}{2\alpha+\sqrt{3}}},
\end{equation}
where $t\leq t_{\rm c}$ with $t_{\rm c}$ being the critical time at which $a\rightarrow\infty$. Note that we are able to obtain big rip even from radiation ($w=\frac{1}{3}$) without introducing a phantom source ($w<-1$).
\end{itemize}

\section{Quiescent universe}

\label{sec:anisocossec}
The quiescent universe which possesses an isotropic initial singularity instead of a chaotic initial state~\cite{Misner:1967uu}, was established by Barrow in 1978~\cite{Barrow1978}, and subsequently studied in detail by many researchers in the field (see~\cite{Goode:1985ab,Hohn:2008zz} and references therein). The model postulates that the early Universe originates from smoothness and regularity and evolves to a state of disorder due to gravitational attraction dominant on large scales. In what follows, we will introduce anisotropy into the scale-independent EMSG and investigate the model in the framework of quiescent cosmology.

We consider the simplest anisotropic generalization of the spatially flat RW spacetime metric, viz., the Bianchi type-I spacetime metric;
\begin{equation}
{\rm d}s^{2}=-{\rm d}t^{2}+A^{2}(t){\rm d}x^{2}+B^{2}(t){\rm d}y^{2}+C^{2}(t){\rm d}z^{2},  \label{aniso}
\end{equation}%
where $\{A(t)$, $B(t), C(t)\}$ are the expansion scale factors along the principal axes $\{x,y,z\}$~\cite{Collins:1972tf,Ellis:1998ct,GEllisBook}. The corresponding average expansion scale factor is $s=(ABC)^{\frac{1}{3}}$, and from which we have $\frac{\dot{s}}{s}=\frac{1}{3}(\frac{\dot{A}}{A}+\frac{\dot{B}}{B}+\frac{\dot{C}}{C})$. In this case, the modified field equations for the isotropic pressure fluid given in~\eqref{em} read
\begin{align}
\frac{\dot{A}\dot{B}}{AB}+\frac{\dot{B}\dot{C}}{BC}+\frac{\dot{C}\dot{A}}{CA}
&=\kappa\rho_{\rm tot},     \label{aniso-friedmann}\\
-\frac{\dot{B}\dot{C}}{BC}-\frac{\ddot{B}}{B}-\frac{\ddot{C}}{C}&=\kappa p_{\rm tot}, \label{aniso-pres1}\\
-\frac{\dot{C}\dot{A}}{CA}-\frac{\ddot{C}}{C}-\frac{\ddot{A}}{A}&=\kappa p_{\rm tot}, \label{aniso-pres2}\\
-\frac{\dot{A}\dot{B}}{AB}-\frac{\ddot{A}}{A}-\frac{\ddot{B}}{B}&=\kappa p_{\rm tot} \label{aniso-pres3},
\end{align}
 where $\rho_{\rm tot}$ (total energy density) and $p_{\rm tot}$ (total pressure) are already defined in Eqs.~\eqref{effdens} and~\eqref{effpres}. Let us define the following total EoS parameter;
\begin{equation}
    w_{\rm tot}\equiv\frac{p_{\rm tot}}{\rho_{\rm tot}}=w+\frac{\alpha (1-w^2)}{4 \alpha w+\sqrt{1+3w^2}}.
\end{equation}
Here we certainly have $w_{\rm tot}=w$ in the absence of EMSG modification ($\alpha=0$). With the use of the definition above, the continuity equation~\eqref{conteqn} in terms of the average scale factor $s$ reads
\begin{equation}
\dot{\rho}+3 \frac{\dot{s}}{s} \rho (1+w_{\rm tot})=0,
\end{equation}
 and the behavior of the energy density becomes
\begin{equation}  \label{aniso-dens}
 \rho \propto s^{-3 (1+w_{\rm tot})}. 
 \end{equation}

 The expansion anisotropy could be quantified through the shear scalar $\sigma^2 \equiv \frac{1}{2}\sigma_{\alpha \beta} \, \sigma^{\alpha \beta}$, where $\sigma_{\alpha \beta} = \frac {1} {2} (u_{\mu;\nu}+u_{\nu; \mu})h^{\mu}_{\;\alpha}h^{\nu}_{\;\beta} - \frac{1}{3} u^{\mu}_{\; ;\mu} \, h_{\alpha \beta}$ is the shear tensor. Here $h_{\mu \nu} = g_{\mu \nu} + u_{\mu} u_{\nu}$ is the so called projection tensor. Using the definitions above along with the modified field equations~\eqref{aniso-friedmann}-~\eqref{aniso-pres3}, for the Bianchi type-I spacetime metric \eqref{aniso}, the shear propagation equation reads
\begin{equation} \label{shearpropagation}
\dot{\sigma}+3\frac{\dot{s}}{s} \sigma=0,  
\end{equation}
and hence, implies
\begin{equation}\label{shear}
\sigma^2 \propto s^{-6}.
\end{equation}
Also, the modified Friedmann equation~\eqref{aniso-friedmann} takes the following form
\begin{equation} \label{aveHubble}
    \frac{{\dot s}^2}{s^2}-\frac{\sigma^2}{3}=\frac{\kappa}{3}\rho_{\rm tot}.
\end{equation}
 Comparing~\eqref{aniso-dens} and~\eqref{shear}, we see that the condition for Barrow's quiescent universe reads $w_{\rm tot}=1$. Note that when  $w_{\rm tot}<1$, the dynamics will approach the vacuum Kasner metric~\cite{Stephani2003} as $s\rightarrow0$. The Kasner metric \cite{Stephani2003} is the exact solution of GR in vacuum for the Bianchi type-I spacetime metric~\eqref{aniso} and represented by the line element ${\rm d}s^2=-{\rm d}t^2+t^{2 p_1}\,{\rm d}x^2+t^{2 p_2}\,{\rm d}y^2+t^{2 p_3}\,{\rm d}z^2$ where $p_1$, $p_2$ and $p_3$ are constants that satisfy the conditions $p_1+ p_2+p_3=p_1^{2}+ p_2^{2}+ p_3^{2}=1$. Since the shear scalar dominates at early times for $w_{\rm tot}<1$, the average scale factor changes as $s \propto t^{\frac{1}{3}}$. In the anisotropic model under consideration, the condition $w_{\rm tot}=1$ in general implies
\begin{equation}
\alpha=\frac{\sqrt{1+3w^2}}{1-3w},
\end{equation}
which leads to
\begin{equation}
    \frac{{\dot s}^2}{s^2}=\frac{\kappa}{3}\left(1+\frac{4 w}{1-3 w}\right)\rho+\frac{\sigma^2}{3},
\end{equation}
except for $w=\{-1,\frac{1}{3},1\}$.  We note that in these limits the material source positively contributes to the modified Friedmann equation (i.e., $\rho_{\rm tot}>0$) when $-1<w<\frac{1}{3}$. Let us now discuss the three special cases: (i) In the case of stiff fluid ($w=1$), we recover the usual quiescent universe independently of the value of $\alpha$, i.e., $\rho \propto s^{-6}$, yet the total energy density reads $\rho_{\rm tot}=\rho(1+2\alpha)$. For the stiff fluid, we recall the condition $\alpha > -\frac{1}{2}$ if $\rho_{\rm tot}>0$. (ii) When the vacuum energy $(w=-1)$ is considered, the total energy density is constant, that is $\rho_{\rm tot}=(1-2 \alpha) \rho_{\rm vac}$, as in GR. Note that for $\alpha=\frac{1}{2}$, we obtain Kasner vacuum solution of GR, in spite of that the universe is filled with the vacuum energy. (iii) In the case of radiation, $w=\frac{1}{3}$,  the total EoS parameter approaches to unity, $w_{\rm tot}\rightarrow1$, as $\alpha\rightarrow\infty$. 

 We would like to point out that in GR, isotropic initial singularity can be achieved only in the presence of stiff fluid described by $w = 1$ that corresponds to the causality limit, and therefore the only possible scenario is Barrow's quiescent universe. However, in EMSG we obtain the total EoS parameter, $w_{\rm tot}$, not by a physical source but by the contribution of an effective source arising from EMSG modification. Thus, our theory permits an ultra-stiff ekpyrotic EoS parameter ($w>1$ in GR) effectively and relaxes the condition for an isotropic initial singularity to $w_{\rm tot} \geq 1$. In such a case, the total energy density dominates as the average scale factor goes to small values (i.e., in the early Universe), hence isotropizes the universe such that it becomes well approximated by RW metric \cite{Ganguly:2021pke}. For instance, when $\alpha<-\frac{\sqrt{3}}{2}$, radiation varies faster than the shear scalar (i.e. $w_{\rm tot} >1$), yet contributes to the modified Friedmann equation negatively (see Eqs.~\eqref{effdens} and \eqref{aveHubble}). If we consider the presence of dust, we have $w_{\rm tot}\geq1$ for $\alpha\geq1$. 

The shear scalar that propagates as exactly as the stiff-fluid does is specific to the Bianchi type-I metric which does not induce any restoring forcelike term in the shear propagation equation (see Eq.~\eqref{shearpropagation}). More complicated spacetimes, however, brings in such terms from the anisotropic spatial curvature of the metric itself \cite{GEllisBook,Ellis:1998ct,Barrow:1997sy}, and hence, renders the shear scalar grow slower than $s^{-6}$ with decreasing $s$. For instance, the Bianchi VII$_0$ metric -- the most general spatially homogeneous and flat anisotropic spacetime metric -- leads to, in the general relativistic universes close to isotropy, the shear scalar to scale as $\sigma^2\propto s^{-5}$ during the dust era~\cite{Barrow:1997sy}. Therefore, the discussion in this section can be extended to more generic anisotropic spacetimes and the resulting shear scalar propagation may further relax the condition required for isotropizing the initial singularity.

\section{Conclusion}
We have considered \textit{the scale-independent energy--momentum squared gravity} (EMSG)~\cite{Akarsu:2018aro} on theoretical ground in order to study some historical and novel cosmological models that it gives rise to. This model provided by $f(T_{\mu \nu }T^{\mu \nu }) = \alpha \sqrt{T_{\mu \nu }T^{\mu \nu }}$ (where $\alpha$ is the dimensionless constant parameter) distinguishes from other EMSG models in the sense that the new terms in the Einstein field equations arising from the scale-independent EMSG theory enter with the same power as the usual terms from Einstein-Hilbert part of the action, thereby the scale-independent EMSG modification affects the field equations independently of the energy density scale considered. Nevertheless, the model violates the energy-momentum tensor (EMT) conservation and matter-current conservation in general and hence, permits a process of matter creation/annihilation in an expanding universe on cosmological scales (in the space between galaxies). Consequently, the continuity equation comes with extra terms and the scale factor dependencies of the energy densities are modified by the model parameter $\alpha$. In this study, we have revisited some historical cosmological models such as (Einstein) static universe and de Sitter/steady state universes, all compatible with the perfect cosmological principle. We have obtained a static, but spatially flat universe when the source is positive vacuum energy ($w=-1$) or stiff fluid ($w=1$). On the other hand, in the presence of a fluid with an EoS parameter in the range $-1<w<1$, we have reproduced the Einstein's original static eternal universe which has a positive (spherical) spatial curvature and a finite size, even without invoking a cosmological constant. In the presence of positive vacuum energy, for $\alpha>\frac{1}{2}$, the scale-independent EMSG drives de Sitter (exponential) expansion as in GR. While a negative vacuum energy can produce only anti-de Sitter expansion in GR, interestingly, in the scale-independent EMSG, de Sitter expansion can also be driven by negative vacuum energy when $\alpha<\frac{1}{2}$. This property of the model is of great importance, since theoretical considerations from high energy physics is indeed in favor of AdS vacua rather than dS vacua. The scale-independent EMSG reproduces the original steady state universe in the presence of dust provided that $\alpha=-1$, and in this case, the effective EMT which comprises the new contributions coming from the modification is equivalent to the creation field proposed by Hoyle~\cite{Hoyle:1948zz} when constructing the original steady state universe model. Also, it provides us with the opportunity to extend the steady state model from dust ($w=0$) to fluids with arbitrary constant EoS parameters. In order to achieve exponential (de Sitter) expansion, the EoS parameter is required to be in the interval $-\frac{1}{3}<w<1$ and also controls the nonarbitrary parameter $\alpha$. We have discussed viable solutions/scenarios for dust ($w=0$) dominated  and radiation ($w=\frac{1}{3}$) dominated universes. In the scale-independent EMSG, dust behaves like a barotropic perfect fluid with constant EoS parameter $\alpha$. As we have mentioned above, dust dominated universe corresponds to a steady state model when $\alpha=-1$. For larger values of $\alpha$, the model leads to accelerated, decelerated and linear (zero acceleration) expansions. As an interesting feature of the scale-independent EMSG, for smaller values of $\alpha$, dust dominated universe ends in a big rip that can be realized solely by a phantom ($w<-1$) source in GR. Similarly, radiation dominated universe corresponds to a steady state model when $\alpha=-\frac{\sqrt{3}}{3}$. For smaller values of $\alpha$, radiation dominated universe ends in a big rip whereas for larger values, exhibits power-law accelerated/decelerated and linear (zero acceleration) expansions. We have also investigated the possible effects of the simplest anisotropic generalization of the spatially flat RW spacetime, viz., the Bianchi type-I spacetime, on the theory. Considering the overall contributions of matter terms to the generalized Friedmann equation as that of a perfect fluid with total EoS parameter $w_{\rm tot}$ (accordingly, with total energy density $\rho_{\rm tot}$ ), we have found that the condition for Barrow's original quiescent universe model is $w_{\rm tot}=1$. We have recovered Barrow's model in the presence of stiff fluid provided that $\alpha>-\frac{1}{2}$, additionally, EMSG could also achieve this scenario for other fluids, particularly for dust (when $\alpha=1$). Furthermore, we have shown that the condition for an isotropic initial singularity relaxes to $w_{\rm tot}>1$ owing to that EMSG effectively allows ultra-stiff EoS parameters ($w>1$) leading to the total energy density, $\rho_{\rm tot}$, grow faster than the expansion anisotropy as we go to the early universe. Likewise, if we switch from Bianchi type-I metric to more generic ones, it is possible to further relax the condition for isotropic initial singularity due to the altered evolution of expansion anisotropy.

We see that both the historical and novel models presented in the current paper are promising enough to justify further investigation of the scale-independent EMSG. Also, in order to address the current tensions such as $H_0$ tension in cosmology, revisiting such historical models in different contexts might enable us to develop new approaches that are hard to come to mind when we strictly stay within the framework of the standard $\Lambda$CDM model assuming GR. We would like to emphasize that we have confined this study to only the cases in which the total contribution of the material stresses to the Friedmann equation, $\rho_{\rm tot}$, is positive since in contrast to GR, otherwise is also possible for EMSG in the presence of spatial curvature or anisotropy. Therefore, with the inclusion of curvature/anisotropy in EMSG, the total energy density can be allowed to take negative values leading to a phantom-like ($w<-1$) behavior. In the light of discussions conducted in this study, it is conceivable that such generalizations of the model will pave the way for constructing some interesting cosmological models, regarding especially the initial and final states of the Universe, which cannot be realized in GR with the use of standard sources. We reserve this analysis for our future works.

\begin{acknowledgments}
 \"{O}.A. acknowledges the support by the Turkish Academy of Sciences in the scheme of the Outstanding Young Scientist Award  (T\"{U}BA-GEB\.{I}P), and the COST Action CA21136 (CosmoVerse). N.M.U. is supported by Boğazi\c ci University Research Fund Grant Number 18541P. 
\end{acknowledgments}

\end{document}